\pdfoutput=1

\documentclass[11pt,twoside,a4paper,cmspaper,final,collab]{cms-tdr}
\def\svnVersion{449948P}\def\cmsCernNoTag{CERN-EP-2017-279}\def\cmsCernDate{\today}\def\cmsMessage{Published in the Journal of High Energy Physics as \href{http://dx.doi.org/10.1007/JHEP03(2018)003}{\doi{10.1007/JHEP03(2018)003.}}}

\begin{document}\cmsNoteHeader{B2G-16-023}

\hyphenation{had-ron-i-za-tion}
\hyphenation{cal-or-i-me-ter}
\hyphenation{de-vices}
\RCS$Revision: 444322 $
\RCS$HeadURL: svn+ssh://svn.cern.ch/reps/tdr2/papers/B2G-16-023/trunk/B2G-16-023.tex $
\RCS$Id: B2G-16-023.tex 444322 2018-02-05 09:41:29Z hirosky $
\newlength\cmsFigWidth
\ifthenelse{\boolean{cms@external}}{\setlength\cmsFigWidth{0.98\columnwidth}}{\setlength\cmsFigWidth{0.48\textwidth}}
\ifthenelse{\boolean{cms@external}}{\providecommand{\cmsLeft}{top\xspace}}{\providecommand{\cmsLeft}{left\xspace}}
\ifthenelse{\boolean{cms@external}}{\providecommand{\cmsRight}{bottom\xspace}}{\providecommand{\cmsRight}{right\xspace}}
\providecommand{\NA}{\ensuremath{\text{---}}}

\newcommand{\Gbulk}{\ensuremath{{\mathrm{G}_{\text{bulk}}}}\xspace}
\newcommand{\mG}{\ensuremath{m_\mathrm{G}}\xspace}
\newcommand{\MPl}{\ensuremath{M_\mathrm{Pl}}\xspace}
\newcommand{\redMPl}{\ensuremath{\overline{M}_\mathrm{Pl}}\xspace}
\newcommand{\resX}{\ensuremath{\cmsSymbolFace{X}}\xspace}
\newcommand{\mX}{\ensuremath{m_{\resX}}\xspace}
\providecommand{\mT}{\ensuremath{m_\mathrm{T}}\xspace}
\newcommand{\mZZ}{\ensuremath{m_{\cPZ\cPZ}}\xspace}
\newcommand{\Zjets}{\ensuremath{\cPZ{+}\text{jets}}\xspace}
\newcommand{\Wjets}{\ensuremath{\PW{+}\text{jets}}\xspace}
\newcommand{\gamjets}{\ensuremath{\gamma{+}\text{jets}}\xspace}
\newcommand{\ptZ}{\ensuremath{\pt^\cPZ}\xspace}
\newcommand{\ptvecZ}{\ensuremath{\ptvec^{\cPZ}}\xspace}

\cmsNoteHeader{B2G-16-023}
\title{Search for ZZ resonances in the $2\ell 2\nu$ final state in proton-proton collisions at 13\TeV}

\date{\today}

\abstract{
A search for heavy resonances decaying to a pair of Z bosons is performed
using data collected with the CMS detector at the LHC.  Events
are selected by requiring two oppositely charged leptons (electrons or muons),
consistent with the decay of a Z boson, and large missing transverse momentum,
which is interpreted as arising from the decay of a second Z boson to
two neutrinos.
The analysis uses data from proton-proton collisions at a center-of-mass
energy of 13\TeV, corresponding to an integrated
luminosity of 35.9\fbinv.
The hypothesis of a spin-2
bulk graviton (\resX) decaying to a pair of Z bosons
is examined for $600\le\mX\le 2500$\GeV and upper limits
at 95\% confidence level are set on the product of the production
cross section and branching fraction of
$\resX\to \cPZ\cPZ$ ranging from 100 to 4\unit{fb}.
For bulk graviton models characterized by a
curvature scale parameter $\tilde{k}=0.5$ in the extra dimension,
the region $\mX < 800$\GeV is excluded, providing the most
stringent limit reported to date.  Variations of the model
considering the possibility of a wide resonance produced exclusively
via gluon--gluon fusion or \qqbar
annihilation are also examined.
}

\hypersetup{
pdfauthor={CMS Collaboration},
pdftitle={Search for diboson resonances in the 2l2nu final state},
pdfsubject={CMS},
pdfkeywords={CMS, physics, diboson resonances}}

\maketitle
\section{Introduction}

The standard model (SM) of particle physics has successfully described
a wide range of high energy phenomena investigated over the decades.
The discovery of a particle compatible with
SM predictions for the Higgs boson~\cite{PhysRevLett.13.321,HIGGS1964132,PhysRevLett.13.508,PhysRevLett.13.585,PhysRev.145.1156,PhysRev.155.1554}
by the ATLAS and CMS
experiments~\cite{ATLASHiggsDiscovery,Chatrchyan201230,CMSHiggsDiscoveryLong}
at the CERN LHC marks an important milestone in the history of particle
physics, providing substantive verification of the SM.
However, the SM lacks a natural means to accommodate the large
hierarchy between gravity and electroweak (EW) scales.
Large loop corrections are necessary to stabilize the SM Higgs
boson mass at the EW scale.  One possible interpretation is that
the measured Higgs boson mass is the result of fine-tuned constants of nature
within the SM.  Alternatively, new physics at the \TeV scale can be invoked
to stabilize the mass of the Higgs boson far below the Planck scale
($\MPl \approx 10^{19}$\GeV).
The spontaneous breaking of EW symmetry in the SM has also
been associated with new dynamics appearing at the \TeV scale.  Examples of
theoretical extensions include the description of a new strongly
interacting sector~\cite{Weinberg:1975gm,Weinberg:1979bn,Susskind:1978ms} or
the introduction of a composite Higgs
boson~\cite{Kaplan:1983fs,Contino:2006nn,Giudice:2007fh}.

Models extending the number of spatial dimensions can also address the
observed difference between the EW and gravitational scales.
A solution postulating the existence of multiple and
potentially large extra spatial dimensions, accessible only for the
propagation of gravity~\cite{ArkaniHamed:1998rs,Antoniadis:1998ig},
was advanced as a way to eliminate the hierarchy between the EW
scale and \MPl .  The model of Randall and
Sundrum~\cite{Randall:1999ee} introduced an alternative
hypothesis, with a single compactified extra dimension and
a modification to the space-time metric by an exponential
``warp'' factor.
Standard model particles reside on a (3+1) dimensional \TeV brane,
while the graviton propagates though the extra dimensional bulk, thereby
generating two effective scales.
These models predict the existence of a tower of massive
Kaluza--Klein (KK) excitations of a spin-2 boson, the KK graviton,
which couples to SM fields at energies on the order of the EW scale.
Such states could be produced at a hadron collider.
However, limits on flavor-changing
neutral currents and EW precision tests place strong constraints
on this model.  The bulk graviton (\Gbulk ) model extends the Randall--Sundrum
model, by addressing the flavor structure of the SM through localization
of fermions in the warped extra
dimension~\cite{Fitzpatrick:2007qr,Antipin:2007pi,Agashe:2007zd},
only confining the Higgs field to the \TeV brane.
The coupling of the graviton to light fermions is highly
suppressed in this scenario and the decays into photons are negligible.
On the other hand, the production of gravitons from gluon--gluon
fusion and their decays into a pair of massive gauge bosons can be
sizable at hadron colliders,
while precision EW and flavor constraints are relaxed to
allow graviton masses in the \TeV range.
The model has two free parameters: the mass of the first mode of the
KK bulk graviton, \mG, and the ratio $\tilde{k}=k/\redMPl$, where $k$ is
the unknown curvature scale of the extra dimension, and
$\redMPl \equiv \MPl / \sqrt{8\pi}$ is the reduced Planck mass.
For values of $\tilde{k}<1$, the width of the KK bulk graviton relative to
its mass is less than $\approx$6\% for \mG as large as 2\TeV, and therefore
a narrow resonance is expected.  Previous direct searches at ATLAS and CMS
have set limits on the cross section for the production of \Gbulk as a
function of \mG~\cite{Aad:2012nev,Aad:2013wxa,Aad:2014xka,Chatrchyan:2012baa,Khachatryan:2014gha,Aaboud:2016okv} using LHC data taken at center-of-mass energies of 7, 8, and 13\TeV.

\begin{figure}[hbtp]
  \centering
    \includegraphics[width=\cmsFigWidth]{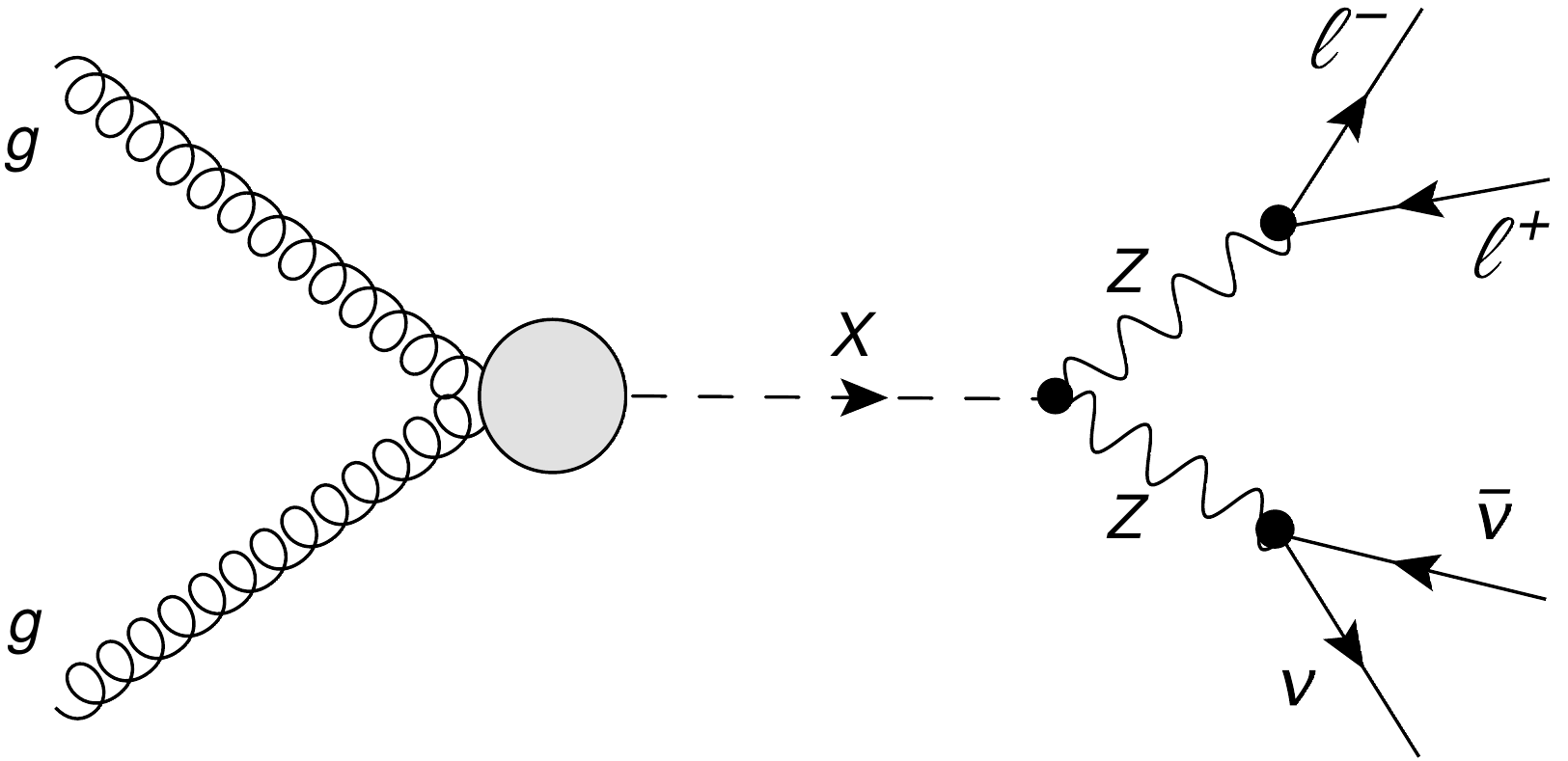}
        \caption{Leading order Feynman diagram for the production of a
        generic resonance \resX via gluon--gluon fusion decaying to the
       \cPZ\cPZ{} final state.}
    \label{fig:Gbulk2ZZ}
\end{figure}

We present a new search for resonances \resX decaying to a pair of \cPZ{} bosons,
in which one of the Z bosons decays into two charged leptons and the other into two
neutrinos $2\ell 2\Pgn$ (where $\ell$ represents either $\Pe$ or $\Pgm$),
as illustrated in Fig.~\ref{fig:Gbulk2ZZ}.
The analysis uses data from proton-proton collisions at a center-of-mass
energy of 13\TeV collected in 2016 and corresponding to an integrated
luminosity of 35.9\fbinv.
The results are compared to expectations for the bulk graviton
model of Refs.~\cite{Fitzpatrick:2007qr,Antipin:2007pi,Agashe:2007zd}.
We also examine variations of the model considering the possibility
of a wide resonance, which is produced exclusively via gluon--gluon fusion
or \qqbar annihilation processes.

The characteristic signature of the $2\ell2\Pgn$ final state includes
two charged leptons with large transverse momenta (\pt) and an overall
imbalance in \pt due to the presence of the undetected
neutrinos.  The imbalance in transverse momentum (\ptvecmiss) is
the negative of the vector sum of the \pt of all
final-state particles; its magnitude is referred to as \ptmiss.
We refer to the observable final states $\Pe\Pe{+}\ptmiss$ and
$\Pgm\Pgm{+}\ptmiss$ as the electron and muon channels, respectively.

The search is performed using the transverse mass (\mT) spectrum of
the two leptons and \ptmiss, where a kinematic edge is expected
from the putative heavy resonance and depends on its invariant mass.
The \mT variable is calculated as:
\begin{equation}
\mT^2 = \left[ \sqrt{(\pt^{\ell\ell})^2 + m^2_{\ell\ell}}
      + \sqrt{(\ptmiss)^2+m^2_{\ell\ell}}\right]^2
      - \left[\vec{p}_{\text{T}}^{\ell\ell}+\ptvecmiss\right]^2,
\label{eqn:MT}
\end{equation}
where $\vec{p}_{\text{T}}^{\ell\ell}\equiv \ptvecZ$ is the \pt of the two lepton
system associated with the leptonic decay of a \cPZ{} boson.
The decay of the second \cPZ{} boson to two invisible neutrinos
is represented by \ptmiss and $m_{\ell\ell}$ in the middle term
provides an estimator of the mass of the invisibly decaying \cPZ{} boson.
This choice has negligible impact on the expected signal at large
\mT, but is found to preferentially suppress backgrounds from \ttbar
and \PW\PW{} decays.

The most significant background to the $2\ell 2\Pgn$ final state is
due to \Zjets production,
where the \cPZ{} boson or recoiling hadrons are not precisely
reconstructed.  This can produce a signal-like final state with \ptmiss
arising primarily from instrumental effects.  Other important sources of
background include the nonresonant production of $\ell\ell$ final states
and \ptmiss, primarily composed of \ttbar and \PW\PW{} production, and the
resonant background from SM production of diboson ({\cPZ}\cPZ{} and {\PW}\cPZ)
events.

Compared to fully reconstructed final states, the branching fraction for the
$2\ell 2\Pgn$ decay mode is approximately a factor of six larger than
that of the four charged-lepton final state, and has less background
than semileptonic channels such as $2\ell$+$2\text{quark}$ ($2\ell 2\PQq$).
For the $2\ell 2\PQq$ channel, the hadronic recoil in the
\Zjets background is kinematically similar to the $2\PQq$ system from
\cPZ{} boson decay.
For events with large \ptmiss, as expected for a
high-mass signal, high \pt jets in the corresponding \Zjets background
are more accurately reconstructed.
This effectively suppresses the background
in the $2\ell 2\Pgn$ channel and the signal purity is  enhanced
relative to the $2\ell 2\PQq$ channel.

\section{The CMS detector}

The central feature of the CMS detector is a 3.8\unit{T} superconducting
solenoid with a 6\unit{m} internal diameter.
Within the solenoid volume are a silicon pixel and strip tracker,
a lead tungstate crystal electromagnetic calorimeter (ECAL), and a brass and
scintillator hadron calorimeter (HCAL), each composed of a barrel and two endcap
sections. Forward calorimeters extend the pseudorapidity coverage ($\eta$)
provided by the
barrel and endcap detectors. Muons are detected in gas-ionization chambers
embedded in the steel magnetic flux-return yoke outside the solenoid.
Events of interest are selected using a two-tiered trigger
system~\cite{Khachatryan:2016bia}. The first level, composed of custom
hardware processors, uses information from the calorimeters and muon detectors
to select events at a rate of around 100\unit{kHz} within a time interval of
less than 4\mus. The second level, known as the high-level trigger,
consists of a farm of processors running a version of the full event
reconstruction software optimized for fast processing, and reduces the event
rate to less than 1\unit{kHz} before data storage.
A detailed description of the CMS detector, together with a
definition of the coordinate system used and the relevant kinematic
variables, can be found in Ref.~\cite{Chatrchyan:2008zzk}.

\section{Event selection and reconstruction}

The signal consists of two \cPZ{} bosons, one decaying into a
pair of oppositely charged leptons and the other to two neutrinos, which
escape direct detection. The final state is thus characterized by
a pair of oppositely charged electrons or muons that are isolated from large
deposits of hadronic energy, having an invariant mass consistent with
that of a \cPZ{} boson, and large \ptmiss.
A single-electron or a single-muon trigger has to be satisfied.
Thresholds on the \pt of the leptons are
115\,(50)\GeV in the electron (muon) channel.
Electron events are triggered by clusters of energy depositions
in the ECAL that are matched to
reconstructed tracks within a range
$\abs{\eta}< 2.5$. Cluster shape requirements, as well as isolation criteria
based on calorimetric and track information, are also applied.
An additional sample of photon plus jet(s) (\gamjets) events is collected for
background modeling based on control samples in data
and is discussed below.  The photon trigger is
similar to the electron trigger, except that a veto is applied on the presence
of a matching track.  For muon events the trigger begins with track fitting
in the outer muon spectrometer.  The outer track is used to seed track
reconstruction in the inner tracker and matching inner-outer track
pairs are included in a combined fit that is used to select muon candidates
in a range $\abs{\eta}< 2.4$.

\subsection{Event reconstruction}

The global event reconstruction (also called particle-flow event
reconstruction~\cite{Sirunyan:2017ulk}) consists of reconstructing and
identifying each individual particle with an optimized combination of
all subdetector information. In this process, the identification of
the particle type (photon, electron, muon, charged hadron, neutral hadron)
plays an important role in the determination of the particle direction
and energy. Photons (\eg coming from \Pgpz\ decays or from electron
bremsstrahlung) are identified as ECAL energy clusters not linked to the
extrapolation of any charged particle trajectory to the ECAL. Electrons
(\eg coming from photon conversions in the tracker material or from
\cPqb-hadron semileptonic decays) are identified as a primary charged
particle track and potentially many ECAL energy clusters corresponding to
this track extrapolation to the ECAL and to possible bremsstrahlung photons
emitted along the way through the tracker material. Muons (\eg from
\cPqb-hadron semileptonic decays) are identified as a track in the central
tracker consistent with either a track or several hits in the muon system,
associated with an energy deficit in the calorimeters. Charged hadrons are
identified as charged particle tracks neither identified as electrons, nor
as muons. Finally, neutral hadrons are identified as HCAL energy clusters
not linked to any charged hadron trajectory, or as ECAL and HCAL energy
excesses with respect to the expected charged hadron energy deposit.

The energy of photons is directly obtained from the ECAL measurement,
corrected for zero-suppression effects. The energy of electrons is
determined from a combination of the track momentum at the main interaction
vertex, the corresponding ECAL cluster energy, and the energy sum of all
bremsstrahlung photons attached to the track. The energy of muons is obtained
from the corresponding track momentum. The energy of charged hadrons is
determined from a combination of the track momentum and the corresponding
ECAL and HCAL energy, corrected for zero-suppression effects and for the
response function of the calorimeters to hadronic showers. Finally, the
energy of neutral hadrons is obtained from the corresponding corrected ECAL
and HCAL energy.

Events are required to have at least one reconstructed interaction vertex.
In case of the existence of multiple vertices, the reconstructed vertex
with the largest value of summed physics-object $\pt^2$ is taken to be
the primary $\Pp\Pp$ interaction vertex. The physics objects are the
jets, clustered using the jet finding
algorithm~\cite{Cacciari:2008gp,Cacciari:2011ma} with the tracks
assigned to the vertex as inputs, and the associated missing transverse
momentum, taken as the negative vector sum of the \pt of those jets.

To reduce the electron misidentification rate, we require the
candidates to satisfy additional identification
criteria that are based on the
shape of the electromagnetic shower in the
ECAL~\cite{Khachatryan:2015hwa}.  Electron candidates
within the transition region between the ECAL barrel
and endcap ($1.479 < \abs{\eta} < 1.566$) are rejected,
because instrumental effects degrade the performance of the reconstruction.
Candidates that are identified as coming from photon
conversions in the detector material are removed.
Photon reconstruction uses the same approach as electrons,
except that photon candidates must not have an assigned track or be
identified as a bremsstrahlung photon from an electron~\cite{CMS:EGM-14-001}.

Muon candidate reconstruction at CMS utilizes several standard
algorithms~\cite{Chatrchyan:2012xi}, two of which are employed in this
analysis. In the first, tracks are reconstructed in the muon system and
propagated inward to the tracker. If a matching track is found, a global
fit is performed to hits in both the silicon tracker and the muon system.
In the second, tracks in the silicon tracker are matched with at least
one muon segment in any detector plane of the muon system, but only
silicon tracking data are used to reconstruct the trajectory of the muon.
To improve efficiency for highly boosted events where the separation
between the two muons is small, we require only one muon to satisfy
the global fit requirement. This results in an efficiency improvement
of 4--18\% for identifying \cPZ{} bosons having \pt in the range of 200--1000\GeV.
The muon misidentification rate is reduced by applying additional
identification criteria based on the number of spatial points measured
in the tracker and in the muon system, the fit quality of the muon track,
and its consistency with the event vertex location.

Leptons produced in the decay of \cPZ{} bosons are expected to be isolated
from hadronic activity in the event.  Therefore, an isolation requirement
is applied based on the sum of the momenta of either charged hadron
PF candidates or additional tracks found in a cone of radius
$\Delta R=0.3$ around each electron or muon
candidate, respectively.  The isolation sum is required to be smaller
than $10\%$ of the \pt of the electron or muon.  For each electron,
the mean energy deposit in the isolation cone coming from other
$\Pp\Pp$ collisions in the same bunch crossing, is estimated following
the method described in Ref.~\cite{Khachatryan:2015hwa}, and subtracted
from the isolation sum.   For muon candidates, only charged tracks
associated with the primary vertex are included and
any additional muons found in the isolation cone are removed from this sum to
prevent rejection of a highly boosted \cPZ{} boson decay.

Jets produced by initial state radiation may accompany signal events
and are also expected to arise from background sources.
The jets are reconstructed from all the PF candidates
using the anti-\kt algorithm~\cite{Cacciari:2008gp,Cacciari:2011ma}
with a radius parameter of $R = 0.4$.
Charged hadron candidates that are not associated with the primary vertex are
excluded. Jet energy corrections are derived from the simulation,
and are confirmed with in situ measurements using the energy balance of
dijet, multijet, \gamjets, and leptonically decaying
\Zjets events~\cite{Khachatryan:2016kdb}.

The \ptmiss is calculated from all the PF candidates,
with momentum scale corrections applied to the candidates.

\subsection{Sample selection}

Events are selected if they include a pair of same-flavor,
oppositely charged leptons that pass the identification
and isolation criteria. The leading (subleading) leptons are required to
have $\pt >120\,(35)$\GeV for the electron channel and $\pt>60(20)$\GeV
for the muon channel. Electrons (muons) are required to be reconstructed
in the range $\abs{\eta}<2.5\,(2.4)$.
To suppress backgrounds that do not include
a \cPZ{} boson, the lepton pair is required to have an invariant mass
compatible with the \cPZ{} boson mass~\cite{Patrignani:2016xqp}
$70<m_{\ell\ell}<110$\GeV.  If more than
one such pair is identified, the pair with invariant mass closest to the
\cPZ{} boson is selected.

The signal region (SR) is defined by additionally requiring
that the \pt of the \cPZ{} boson candidate satisfies $\ptZ>100$\GeV,
$\ptmiss>50$\GeV, and the angular difference between $\ptvecZ$
and \ptvecmiss satisfies $\abs{\Delta \phi(\ptvecZ, \ptvecmiss)}>0.5$\unit{radians}.
The SR selection largely suppresses the backgrounds, which are
primarily concentrated at low \ptZ and low \ptmiss.
In the case of a signal we expect two highly boosted \cPZ{} bosons, therefore, the
$|\Delta \phi(\ptvecZ, \ptvecmiss)|$ distribution is correspondingly
peaked around $\pi$ in contrast to a relatively flat distribution in
the \Zjets background where \ptvecmiss arises from instrumental effects.

\section{Signal and background models}

Two versions of the signal model are examined.
For our benchmark model, signal events are generated at
leading order for the bulk graviton model of
Refs.~\cite{Fitzpatrick:2007qr,Antipin:2007pi,Agashe:2007zd} using the
\MGvATNLO~2.3.3 event generator~\cite{Alwall:2014hca}.
Because the expected width is small compared to detector resolution for
reconstructing the signal, we use a zero width
approximation~\cite{Oliveira:2014kla} for generating signal events.
A more general version of the bulk graviton decaying to {\cPZ}\cPZ{} is generated using
JHU~Generator~7.0.2~\cite{Gao:2010qx,Bolognesi:2012mm,Chen:2013waa}.
We model a bulk graviton as in
Refs.~\cite{Chatrchyan:2013mxa,Khachatryan:2014kca} and introduce variable
decay widths up to 30\% of \mX.  Production of the wide resonance via
gluon fusion and \qqbar annihilation
are generated separately.
Generated events are interfaced to {\PYTHIA~8.212}~\cite{Sjostrand:2006za} for
parton showering and hadronization.
The renormalization and factorization scales are set to the resonance mass.
Parton distribution functions (PDFs) are modeled using the
NNPDF~3.0~\cite{Ball:2014uwa} parametrization. Signal samples are
generated in the mass range 600--2500\GeV for each tested model.
We simulate both signal and background using a \GEANTfour-based
model~\cite{Agostinelli:2002hh,Allison:2006ve,Asai:2015xno} of the CMS detector and process the
Monte Carlo (MC) events using the same reconstruction algorithms as for
data.  All MC samples include an overlay of additional minimum bias
events (also called ``pileup''), generated with an approximate
distribution for the number of expected additional $\Pp\Pp$ interactions,
and events are reweighted to match the distribution observed in data.

The largest source of background arises from the production of
\Zjets events, characterized by a transversely boosted \cPZ{} boson and
recoiling hadrons.  The observation of \ptmiss  in these events primarily
results from the mismeasurement of jet or lepton \pt.
While this process may be modeled exclusively using simulated events,
the description of detector instrumental effects can be improved
by constructing a background estimate based on control samples in data.
We use a sample of \gamjets data with a reweighting procedure to reproduce
the kinematics of the \cPZ{} boson in \Zjets events, exploiting the intrinsic
similarity of the recoiling hadrons balancing the \pt of the \cPZ{} boson
or the photon. The procedure also employs a sample of \Zjets events generated using
the {\MADGRAPH}5\_aMC@NLO framework with next-to-leading order (NLO)
matrix elements for final states with up to two additional partons.
The merging scheme of Frederix and Frixione is employed for matching to parton showers using
a merging scale $\mu_Q=30\GeV$~\cite{Frederix:2012ps}.
The
inclusive cross section is recalculated to include next-to-next-to-leading
order (NNLO) QCD and EW corrections from {\FEWZ}~3.1~\cite{Li:2012wna}.
We use the \Zjets differential cross section measurement as a function
of \ptZ in CMS data to reweight each event in the MC
sample at the generator level to match the dependence observed in data.
The differential cross section measured in \gamjets data is first corrected for
backgrounds producing physical \ptmiss, such as \Wjets events.
The reconstructed \gamjets events in data are
then reweighted as a function of $\pt^\gamma$ and $\abs{\eta^\gamma}$
to match the corrected \Zjets spectra in simulation
for electron and muon channels separately.
This procedure transfers the lepton trigger and identification efficiencies
from \Zjets, into the \gamjets data sample.  For calculation of the \mT
variable in Eq.~(\ref{eqn:MT}), the photon is randomly assigned a mass
based on the measured \cPZ{} boson mass distribution as a function of the
\cPZ{} boson \pt.
Finally to account for small energy scale and resolution
differences in the \ptmiss between \gamjets and \Zjets events, we
fit the parallel and perpendicular components of the hadronic recoil relative
to the reconstructed boson in both samples
using a Gaussian model in bins of boson \pt.  The differences are
used to correct the \gamjets data as a function of photon \pt.

The nonresonant backgrounds can be significant in regions of
large \ptmiss  due to the presence of neutrinos in the final state.
A method based on control samples in data is used to more precisely model
this background. The method uses dilepton samples consisting
of \Pe\Pgm\ pairs to describe the expected background
in $\ell\ell$ (\Pe\Pe\ or \Pgm\Pgm) events.  This utilizes the fact
that \Pe\Pgm\ pairs in the nonresonant background have very similar
kinematic behavior and cross sections compared to the $\ell\ell$
final states.  Events with at least one \Pe\Pgm\ pair are selected.
If more than one pair is present, the pair having an
invariant mass closest to that of the \cPZ{} boson is selected.
The normalization of event yields
between $\ell\ell$ and \Pe\Pgm\ events is estimated using events
outside the \cPZ{} boson mass selection window. Because of
effects due to different trigger requirements and identification
efficiencies, variances are observed in the lepton \pt distributions
compared to the single-flavor samples.  Therefore when modeling the
electron (muon) channel, event-based weighting factors are applied to
correct the \pt distribution of the muon (electron) in the \Pe\Pgm\ data
for these observed differences. The trigger efficiency is also applied in the
background sample to simulate the single-lepton trigger efficiency.
The correction corresponding to either the electron or muon channel
is applied based on the \pt and $\abs{\eta}$ of both leptons.

\begin{figure}[hbtp]
  \centering
    \includegraphics[width=\cmsFigWidth]{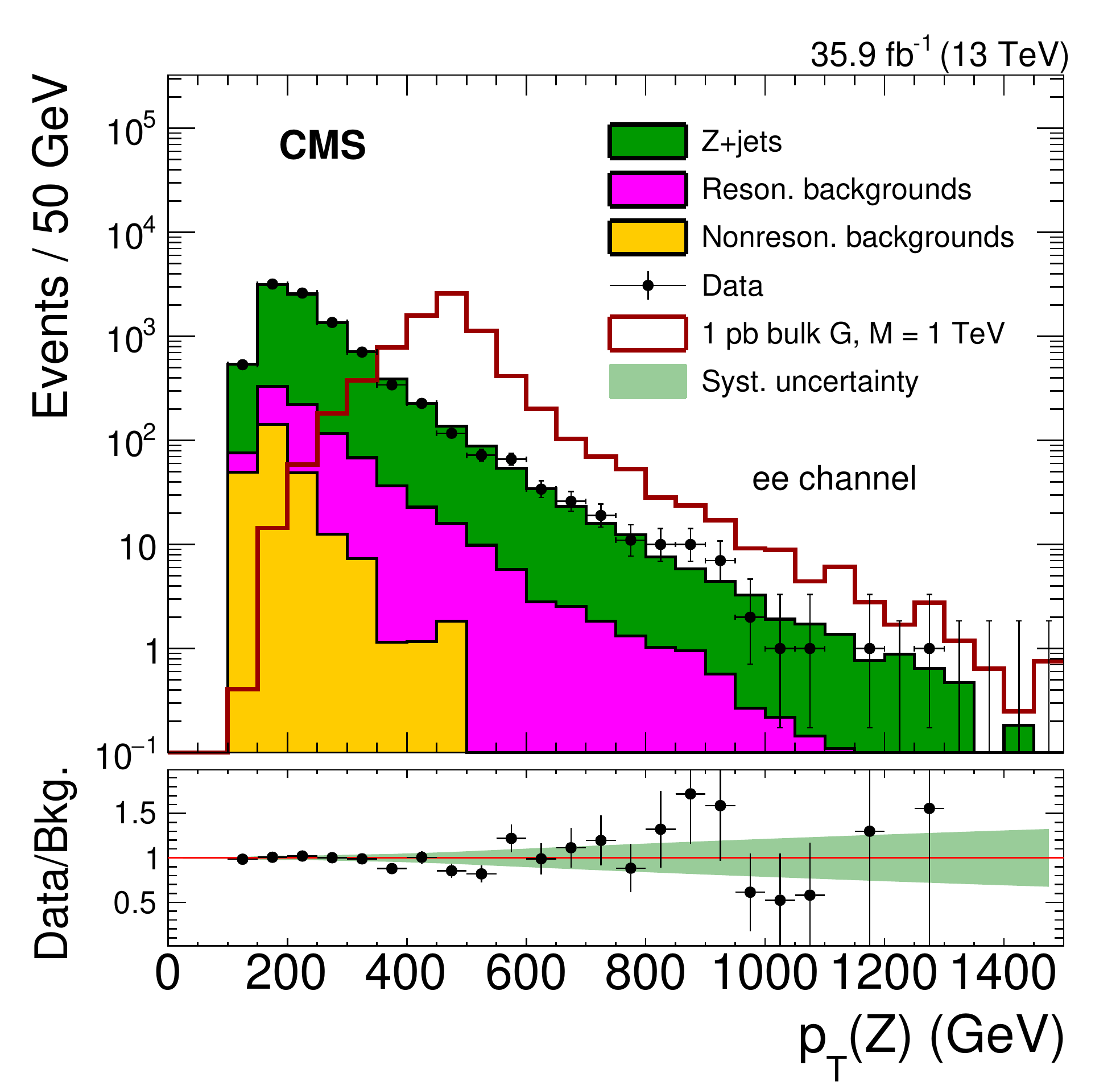}
    \includegraphics[width=\cmsFigWidth]{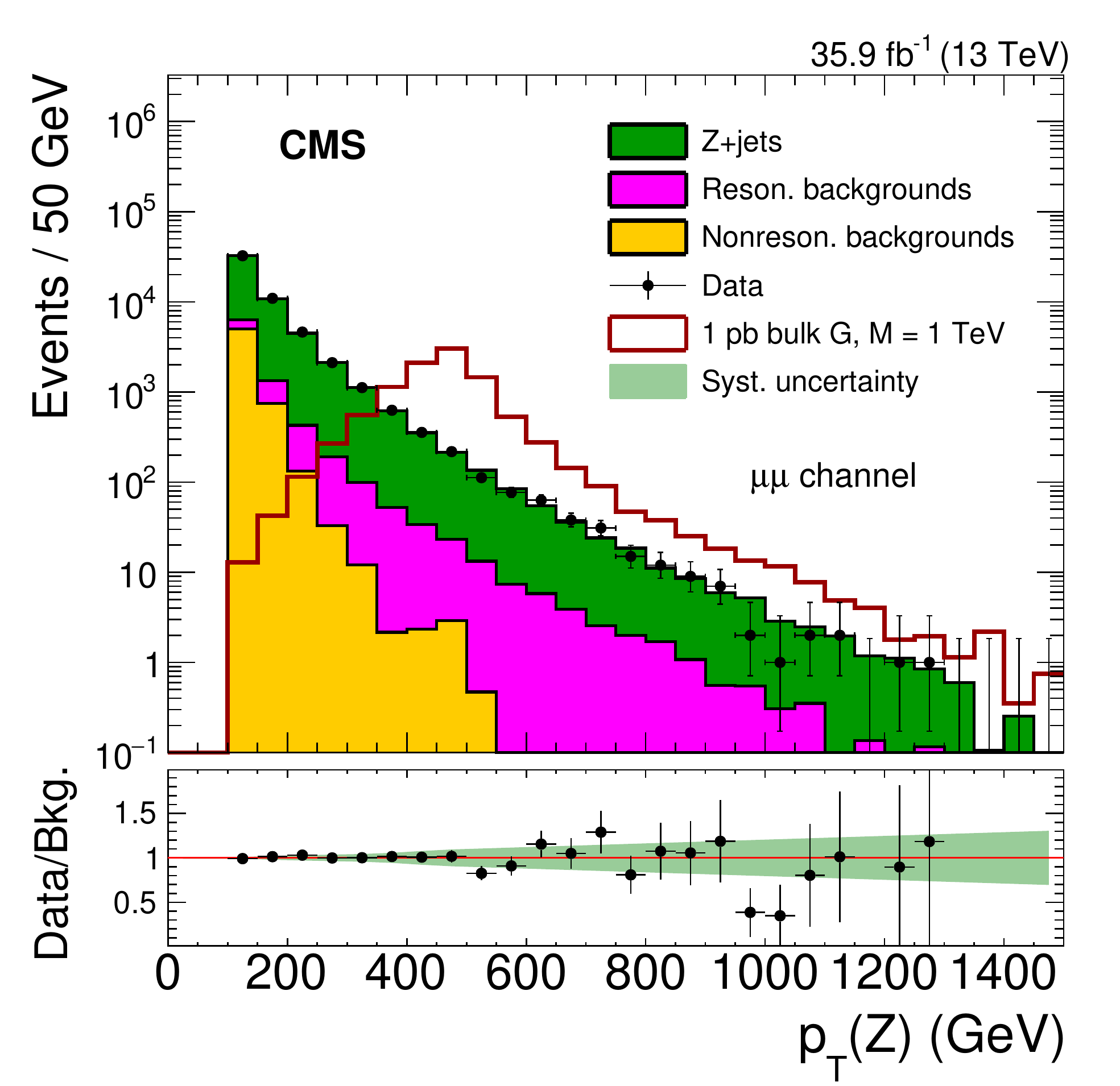}
    \caption{The $\pt^{\cPZ}$ distributions for electron (left)
    and muon (right) channels comparing the data and
    background model based on control samples in data.
    The lower panels give the ratio of data to the
    prediction for the background.  The shaded band shows
    the systematic uncertainties in background,
    while the statistical uncertainty in the data is shown by the error bars.
     The expected distribution for a zero width bulk graviton resonance
    with a mass of 1\TeV is also shown
    for a value of 1\unit{pb} for the product of cross section and branching fraction
     $\sigma(\Pp\Pp \to \resX\to \cPZ\cPZ)\, \mathcal{B} (\cPZ\cPZ\to2\ell2\Pgn)$.
}
    \label{fig:ZpT}
\end{figure}

The irreducible (resonant) background arises mainly from the SM
$\PQq\Paq \to \cPZ\cPZ{} \to 2\ell 2\Pgn$ process and is
modeled using MC samples generated by \POWHEG~2.0~\cite{Nason:2004rx,Frixione:2007vw},
at NLO in QCD and leading order in EW calculations.
We also apply NNLO QCD~\cite{Grazzini:2015hta} and NLO EW corrections to the
production processes~\cite{Bierweiler:2013dja,Gieseke:2014gka}.
These are applied as a function of \mZZ and on average are
1.11 and 0.95 for the NNLO QCD and NLO EW corrections, respectively.
Smaller contributions from \PW\cPZ{} and {\ttbar}\cPZ{} decays are modeled at
NLO using {\MADGRAPH}5\_aMC@NLO.

\begin{figure}[hbtp]
  \centering
    \includegraphics[width=\cmsFigWidth]{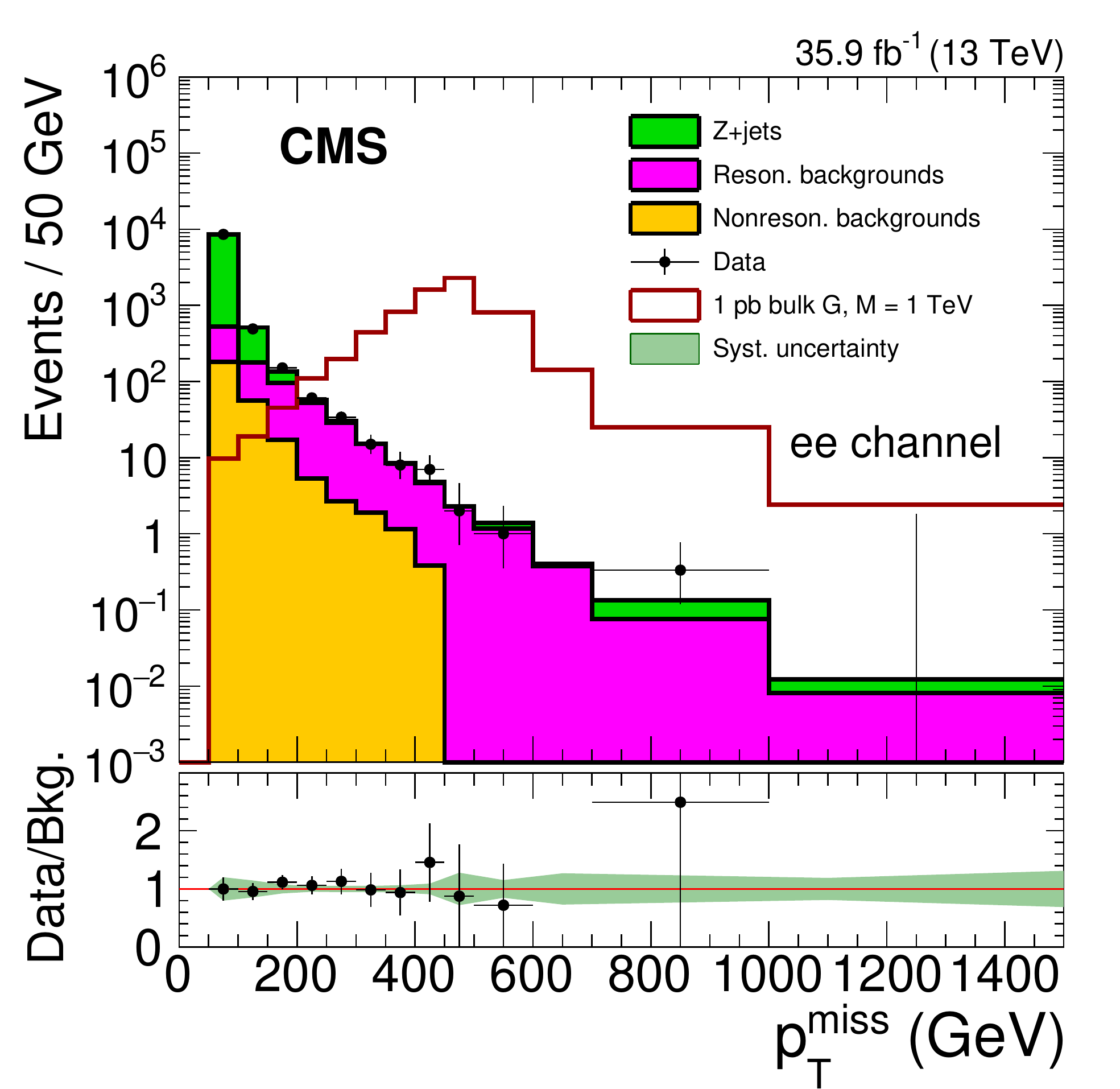}
    \includegraphics[width=\cmsFigWidth]{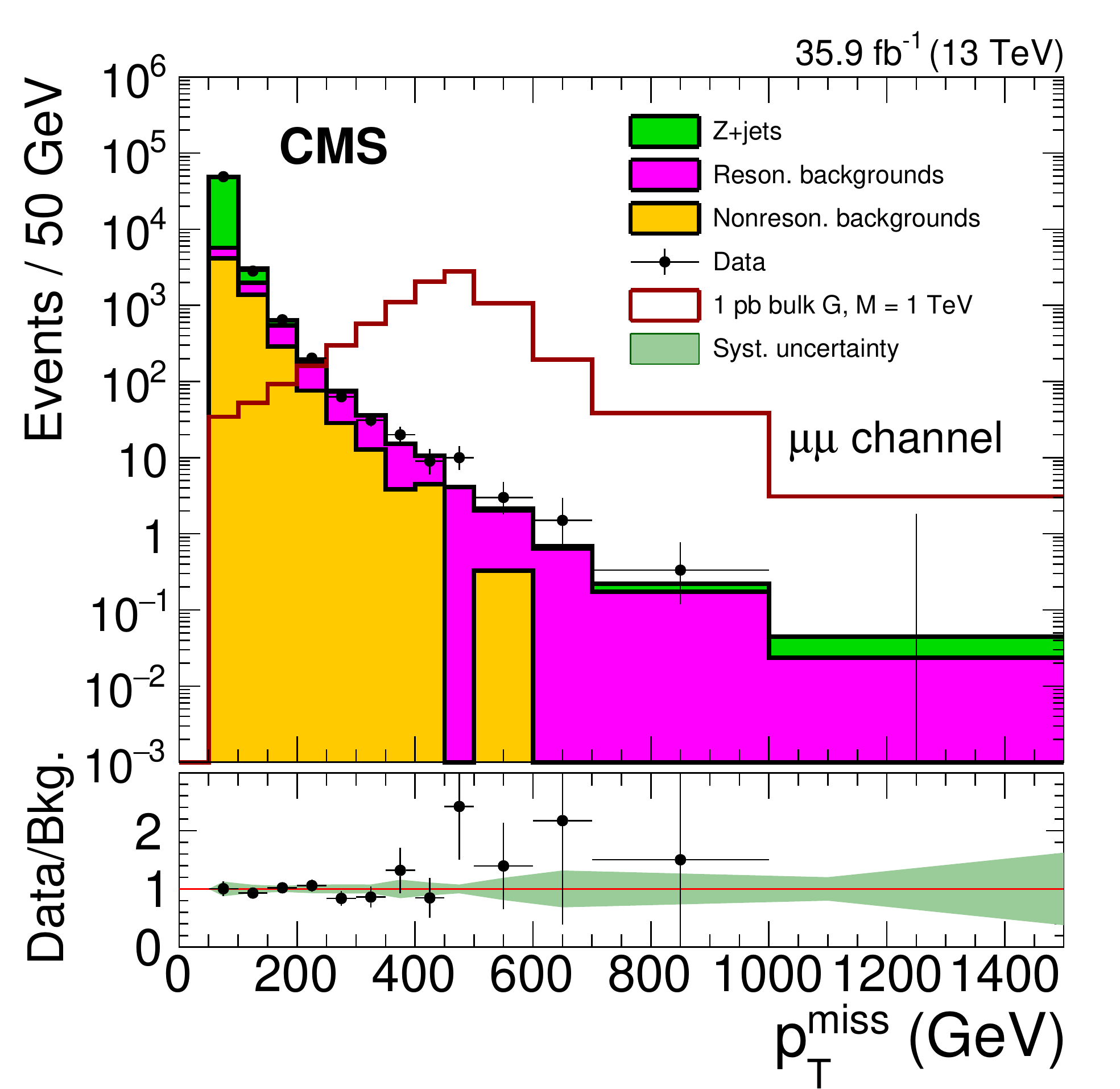}
    \caption{The \ptmiss for electron (left) and muon (right) channels
    comparing the data and background model based on control samples in data.
    The expected distribution for a zero width bulk graviton resonance
    with a mass of 1\TeV is also shown
    for a value of 1\unit{pb} for the product of cross section and branching fraction
       $\sigma(\Pp\Pp \to \resX\to \cPZ\cPZ)\, \mathcal{B} (\cPZ\cPZ\to2\ell2\Pgn)$.
    The lower panels show the ratio of data to the
    prediction for the background.
    The shaded band shows the systematic uncertainties in background,
    while the statistical uncertainty in the data is shown by the error bars.
    }
    \label{fig:MET}
\end{figure}

Figure~\ref{fig:ZpT} shows the comparison of background models and data
for the \pt distribution of the reconstructed \cPZ{} boson after all
corrections are applied.
Figure~\ref{fig:MET} shows the data and background prediction of the
\ptmiss distribution after all corrections are applied.  The \ptmiss is an
essential variable to examine the quality of the background modeling and
the understanding of the systematic uncertainties.  All the systematic
uncertainties are propagated to the \ptmiss distributions and shown
as the uncertainty band on the ratio plots in the lower panels of the figure.
Also shown in Figs.~\ref{fig:ZpT} and~\ref{fig:MET} is the expected
signal distribution assuming a bulk graviton with 1\TeV mass
 and an arbitrary product of the cross section and branching fraction
 $\sigma(\Pp\Pp \to \resX\to \cPZ\cPZ)\, \mathcal{B} (\cPZ\cPZ\to2\ell2\Pgn)$
 of 1\unit{pb}.

\section{Systematic uncertainties\label{sec:systematics}}

Systematic uncertainties can affect both the normalization and differential
distributions of signal and background.  Individual
sources of systematic uncertainties are evaluated
by studying the effects of parameter variations within one standard deviation
relative to their nominal values and propagating the result into the \mT
template distributions that are used to evaluate signal cross section limits.
The various categories of systematic uncertainties affecting these
distributions are described below and summarized in
Table~\ref{tab:unc_summary} for both electron and muon channels.

\begin{table}[htbp]
\topcaption{Summary of the normalization uncertainties that are
included in the statistical procedure for the electron and muon channels.
All values are listed in percentage units and similar categories are
grouped for brevity.  Sources that do not apply or are found to be
negligibly small are marked  ``\NA'' or ``(\NA),'' respectively.
Integrated luminosity and theoretical uncertainties are evaluated separately
for effects on normalizations, while all the other uncertainties are considered
simultaneously with shape variations in the statistical analysis.
Values in the signal column refer to the hypothetical spin-2 bulk
graviton signal with a mass of 1\TeV.
\label{tab:unc_summary}}
\centering
\begin{tabular}{l l c c c c c }
		&	Source	&	Signal	&	{\Zjets}	&	Resonant	&	Nonresonant	\\
		&		&	(\%)	&	(\%)	&	        (\%)	&	        (\%)	\\ \hline
		&	Integrated luminosity	&	2.5	&	2.5	&	2.5	&	2.5	\\
		&	PDF: cross section	&	\NA	&	2.3	&	1.7	&	\NA	\\
		&	Scale: cross section	&	\NA	&	3.5	&	3.0	&	\NA	\\
		&	EW NLO correction	&	\NA	&	\NA	&	3.0	&	\NA	\\[2ex]
	\multirow{9}{*}{\parbox{8ex}{Electron channel}}	&	PDF: acceptance	&	1.0	&	3.4	&	1.0	&	\NA	\\
	    &	Scale: acceptance	&	(\NA)	&	22.7	&	2.9	&	\NA	\\
		&	Trigger/identification eff.	&	2.1	&	\NA	&	0.4	&	\NA	\\
		&	\ptZ reweighting	&	\NA	&	6.8	&	\NA	&	\NA	\\
		&	Nonresonant norm.	&	\NA	&	\NA	&	\NA	&	10.0	\\
		&	\pt /energy scale	&	(\NA)	&	\NA	&	4.6	&	\NA	\\
		&	Jet energy resolution	&	(\NA)	&	\NA	&	6.8	&	\NA	\\
		&	Unclustered energy	&	(\NA)	&	\NA	&	5.5	&	\NA	\\
		&	Hadronic recoil	&	\NA	&	3.4	&	\NA	&	\NA	\\[2ex]
	\multirow{9}{*}{\parbox{8ex}{Muon channel}}	&	PDF: acceptance	&	1.0	&	3.4	&	1.0	&	\NA	\\
	    &	Scale: acceptance	&	(\NA)	&	13.1	&	2.9	&	\NA	\\
		&	Trigger/identification eff.	&	3.6	&	1.0	&	1.0	&	1.0	\\
		&	\ptZ reweighting	&	\NA	&	3.2	&	\NA	&	\NA	\\
		&	Nonresonant norm.	&	\NA	&	\NA	&	\NA	&	2.4	\\
		&	\pt /energy scale	&	(\NA)	&	\NA	&	7.4	&	\NA	\\
		&	Jet energy resolution	&	(\NA)	&	\NA	&	5.6	&	\NA	\\
		&	Unclustered energy	&	(\NA)	&	\NA	&	6.3	&	\NA	\\
		&	Hadronic recoil	&	\NA	&	2.0	&	\NA	&	\NA	\\
\hline
\end{tabular}
\end{table}

Uncertainties from trigger efficiencies, lepton identification and isolation
requirements, and tracking efficiency can affect signal and background
estimates obtained from both simulation and from control samples in data.
The combined effect of these uncertainties on the normalizations
of the various samples is found to be 0.4--3.6\%.

Uncertainties of 6.8\,(3.2)\% for the electron (muon) channel are assigned to the
reweighting procedure for the \Zjets background.
For the nonresonant background, modeling of trigger and
lepton identification
efficiencies relative to the \cPZ{} boson data and the size of the sideband
samples contribute the major uncertainties in the expected event yields.
These are estimated to affect the normalization by 10\,(2.4)\% for the
electron (muon) channel.

The lepton momenta, and photon and jet energies are recalculated by varying
their respective corrections within scale uncertainties.  These uncertainties
affect event selection and the detector response corrected \ptmiss,
contributing a variation of 4.6\,(7.4)\% to the template normalizations for
the MC-generated resonant backgrounds in the electron (muon) channel.
Their corresponding effect on acceptance for the signal is negligible.
The modeling of jet resolution and the
correction applied to unclustered energy are similarly considered for the MC
samples and found to contribute an uncertainty of $\approx$6\% each
to the resonant background normalization.  The effect of variations in
corrections to the modeling of recoil in the \Zjets background
is found to be 3.4\% and 2.0\% for the electron and muon channel,
respectively.

Uncertainties arising from the PDF model and renormalization and factorization
scales in fixed-order calculations affect signal and simulated
backgrounds, modifying predictions for both the production cross-section
and the acceptance.
We estimate the effect of PDF uncertainties by evaluating the
complete set of NNPDF~3.0 PDF eigenvectors, following the
PDF4LHC prescription~\cite{Ball:2014uwa,Butterworth:2015oua}.
This contributes a variation of 1.0--3.4\% to the MC background models.
The production of bulk gravitons is modeled by a fusion process with
gluons having large Bj\"orken-$x$, where parton luminosities are generally
not well-constrained by existing PDF models. The PDF uncertainties in the
signal production cross section depend on \mX and range from 10--50\%,
but modify the acceptance by only about 1\%.

The effect of scale variations is assessed by varying the original
factorization and renormalization scales by factors of 0.5 or 2.0.
The scale uncertainties are estimated to be about 3--3.5\% each in the
production cross section and acceptance for the resonant background.
For the \Zjets background, the scale choice modifies the normalization by 3.5\%.
The acceptance varies by 23~(13)\% in the electron (muon) channel and the corresponding
effect is negligibly small for the signal.  An uncertainty of 3.0\% is
estimated for the (N)NLO correction to the resonant background.
The uncertainty assigned to the integrated luminosity measurement
is 2.5\%~\cite{CMS-PAS-LUM-17-001} and is applied to the signal and simulated
backgrounds.

In the treatment of systematic uncertainties, both normalization
effects, which only alter the overall yields of individual contributions,
as well as shape variations, which also affect their distribution,
are taken into account for each source individually.

\section{Statistical interpretation}

The \mT distribution is used as the sensitive variable to search for a new
resonance decaying to {\cPZ}\cPZ{} with the subsequent decay
$\cPZ\cPZ\to 2\ell2\Pgn$. For both the electron and muon channels,
a binned shape analysis is employed. The expected numbers of background
and signal events scaled by a signal strength modifier are combined to
form a binned likelihood calculated using each bin of the \mT distribution.

\begin{table}[htbp]
\topcaption{Event yields for different background contributions
and those observed in data in the electron and
muon channels. \label{tab:yields}}
\centering
\begin{tabular}{l D{,}{\pm}{-1} D{,}{\pm}{-1}}
{}          	  &  \multicolumn{1}{c}{Electron channel}   & \multicolumn{1}{c}{Muon channel} \\ \hline
Data 		  &   \multicolumn{1}{c}{9336}              &  \multicolumn{1}{c}{52806}          \\[2ex]
Z$+$jets 	          &   8421 , 203    &  44253 , 336 \\
Resonant          &    637 ,  38    &   2599 , 164 \\
Nonresonant	  &    271 ,  28    &   5961 , 211 \\[2ex]
Total background  &   9329 , 208    &  52813 , 439 \\
\hline
\end{tabular}
\end{table}

\begin{figure}[hbtp]
  \centering
    \includegraphics[width=\cmsFigWidth]{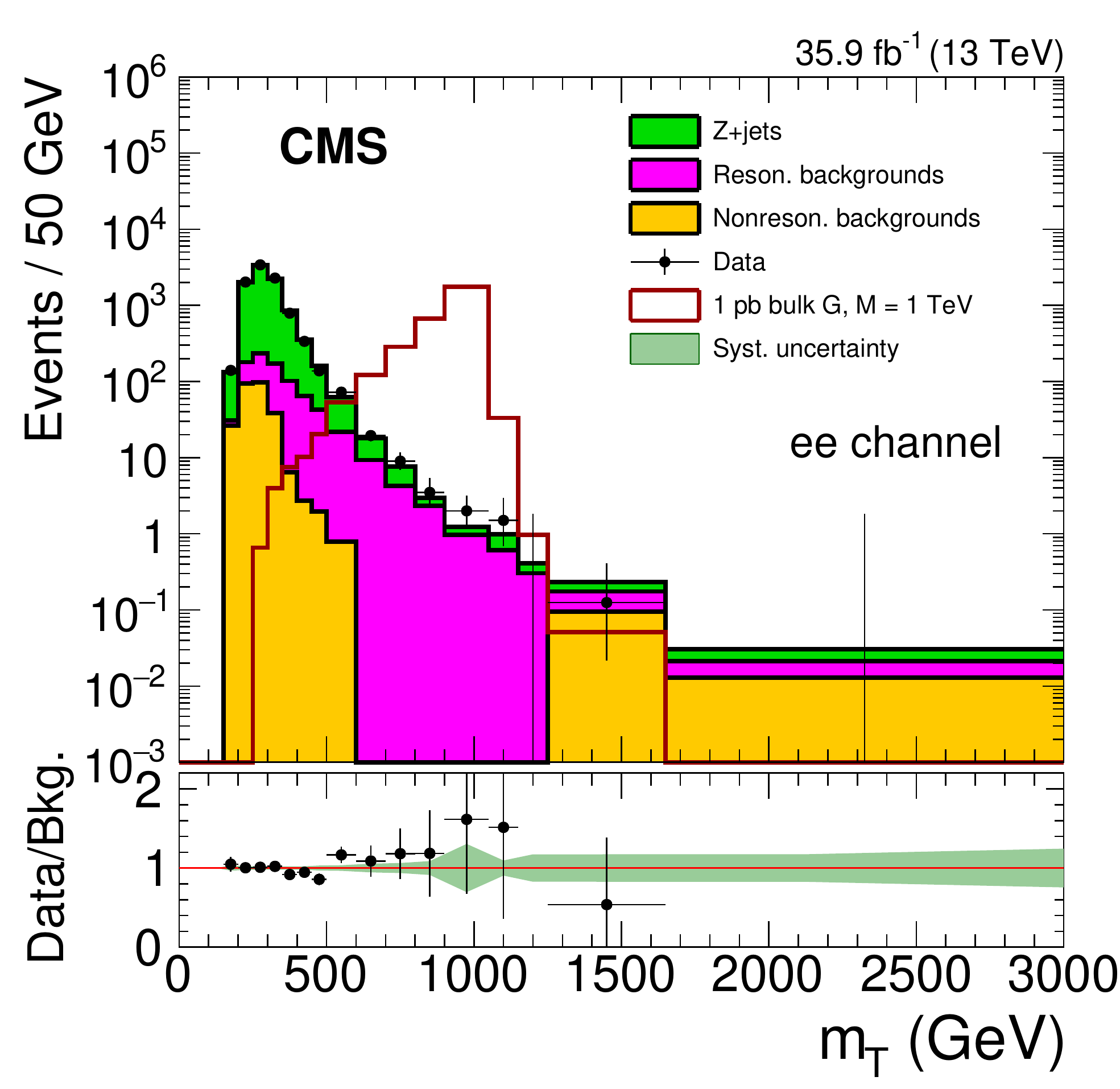}
    \includegraphics[width=\cmsFigWidth]{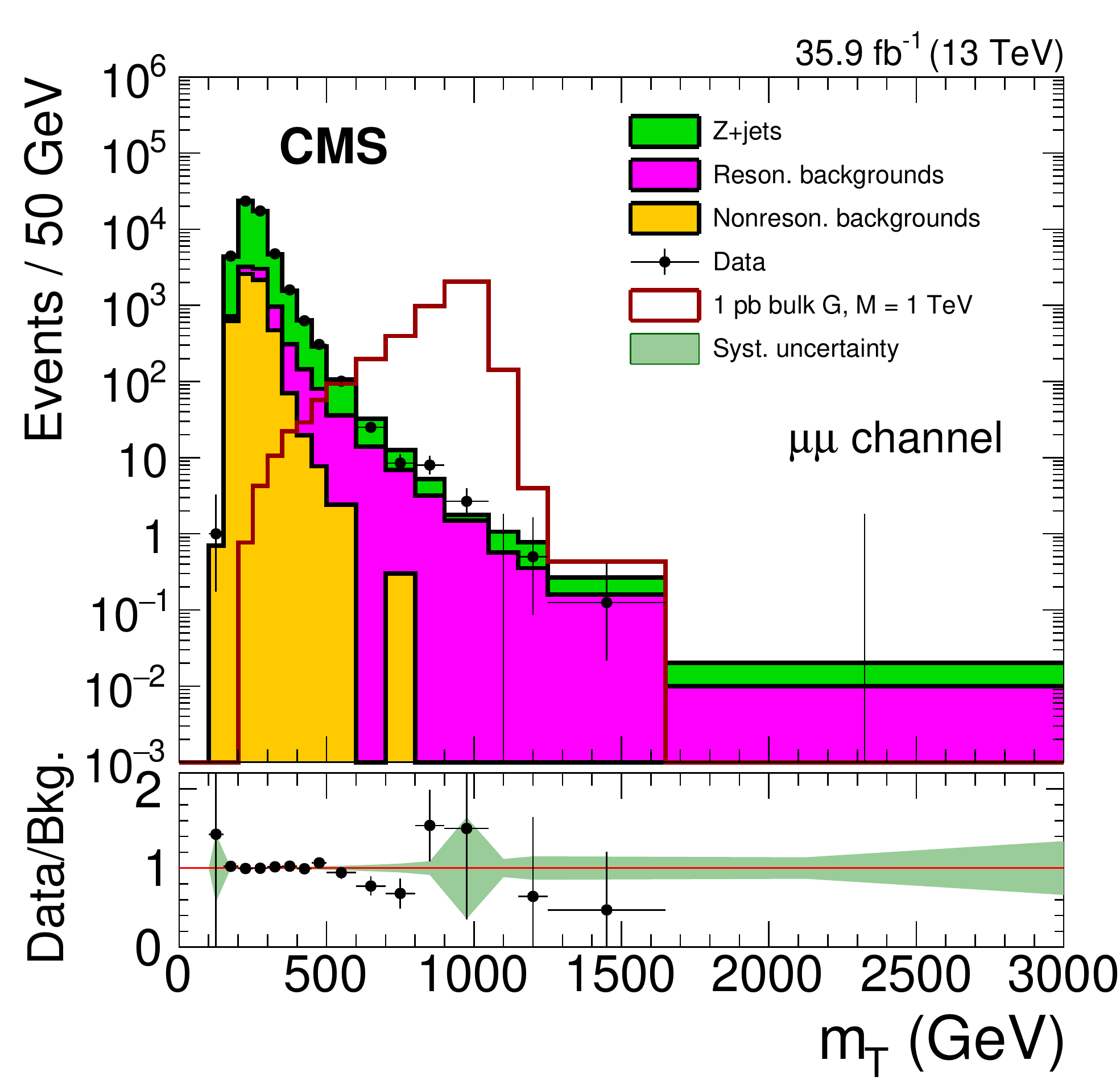}
    \caption{The \mT distributions for electron (left) and muon (right)
    channels comparing the data and background model based on control
    samples in data, after fitting the background-only model to the data.
    The expected distribution for a zero width bulk graviton resonance
    with a mass of 1\TeV is also shown
    for a value of 1\unit{pb} for the product of branching fraction
    and cross section
    $\sigma(\Pp\Pp \to \resX\to \cPZ\cPZ)\, \mathcal{B} (\cPZ\cPZ\to2\ell2\Pgn)$.
    The lower panels show the ratio of data to the prediction for the
    background. The shaded bands show the systematic uncertainties in
    the background,
    while the statistical uncertainty in the data is shown by the error bars.
    }
    \label{fig:MT}
\end{figure}

The results of a simultaneous fit of the predicted backgrounds to data,
combining electron and muon channels, and including the estimated systematic
uncertainties are summarized in Table~\ref{tab:yields}.  Figure~\ref{fig:MT}
shows the post-fit \mT distributions in the SR using only the background
models.  The expected distribution for a bulk graviton signal with
a mass of 1\TeV and
an arbitrary product of cross section and branching fraction
$\sigma(\Pp\Pp \to \resX\to \cPZ\cPZ)\, \mathcal{B} (\cPZ\cPZ\to2\ell2\Pgn)$
of 1\unit{pb} is also shown.
The observed distributions are in agreement with fitted SM background
predictions.

Upper limits on the product of cross section and branching fraction for the resonance
production $\sigma(\Pp\Pp \to \resX \to \PZ\PZ)$
are evaluated using the asymptotic approximation~\cite{Cowan:2010js}
of the modified frequentist approach CL$_\mathrm{s}$~\cite{Junk:1999kv,Read:2002hq,CMS-NOTE-2011-005}.
The same simultaneous combined fit is performed using signal and background
distributions after application of the SR selection, to extract the upper limits
for a given signal hypothesis. Statistical uncertainties in the background
modeling are taken into account by fluctuating the predicted background
histograms within an envelope according to uncertainties in each bin.
Systematic uncertainties are treated as nuisance parameters, constrained with
Gaussian or log-normal probability density functions in the maximum likelihood fit.
For the signal, only uncertainties related to luminosity and acceptance contribute
in the limit setting procedure.
When the likelihoods for electron and muon channels are combined,
the correlation of systematic effects is taken into account.

\section{Results}

\begin{figure}[hbtp]
  \centering
    \includegraphics[width=\cmsFigWidth]{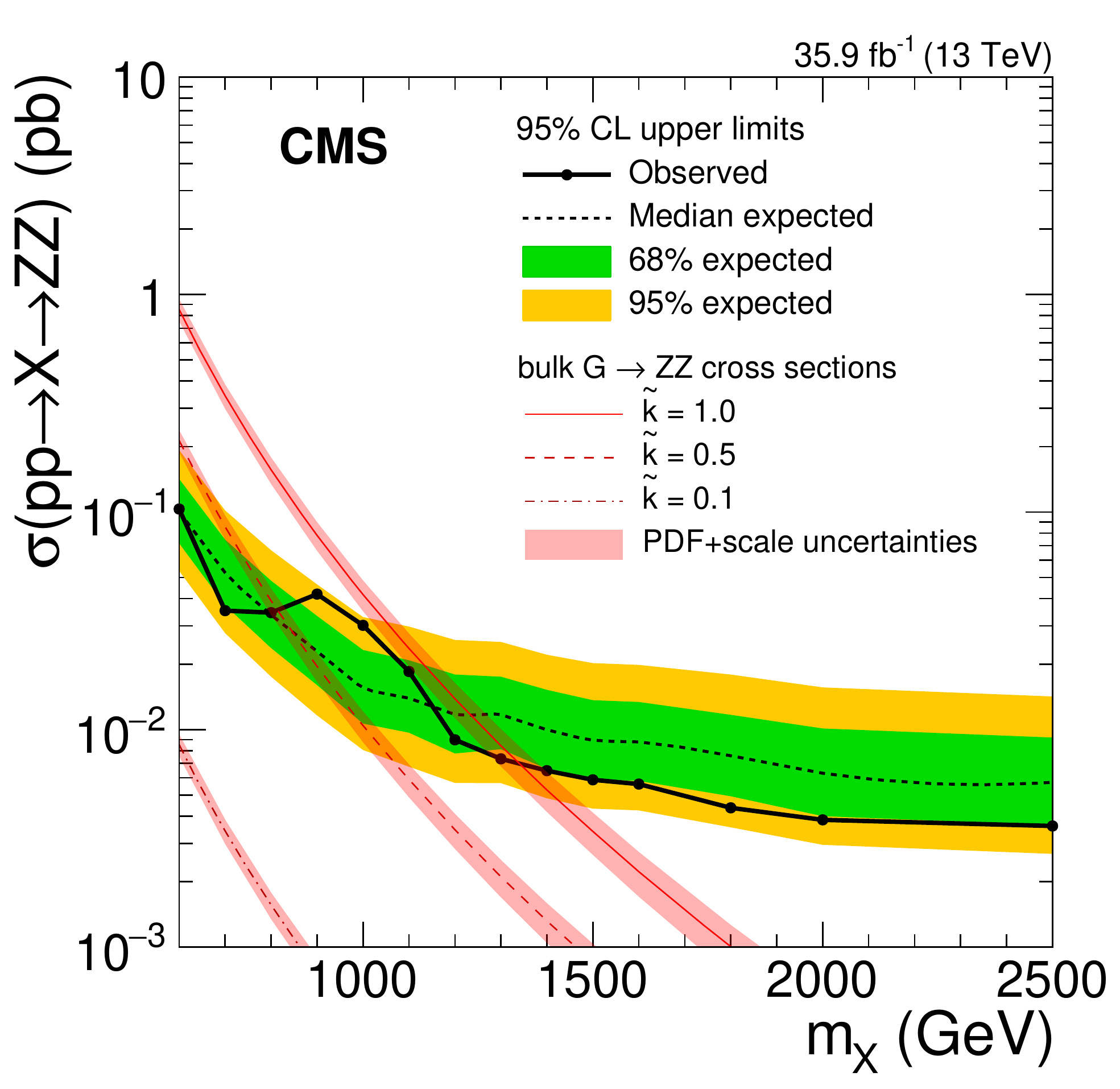}
        \caption{Expected and observed limits on the product of cross section
        and branching fraction of a new spin-2 heavy resonance
        $\resX \to \cPZ\cPZ$, assuming zero
        width, based on the combined analysis of the electron and muon
        channels.  Expectations for the production cross section
        $\sigma(\Pp\Pp \to \resX\to \cPZ\cPZ)$ are also shown for
        the benchmark bulk graviton model for three values of the
        curvature scale parameter $\tilde{k}$. }
    \label{fig:limits_narrow_sr}
\end{figure}

The expected and observed upper limits on the product of the
resonance cross section and the branching fraction for
$\resX \to \cPZ\cPZ$ are determined
at the 95\% confidence level (CL) for the zero width benchmark model as a
function of \mX and shown in Fig.~\ref{fig:limits_narrow_sr} for the
{\Pe}\Pe\ and {\PGm}\PGm\ channels combined.
Expectations for $\sigma(\Pp\Pp \to \resX\to \cPZ\cPZ)$
are also normalized to the calculations of Ref.~\cite{Oliveira:2014kla}
and shown as a function of
the bulk graviton mass for three values of the curvature scale parameter
$\tilde{k}=(1.0, 0.5, 0.1)$. The hypothesis of $\tilde{k}=0.5$ can be
excluded for masses below 800\GeV at 95\% CL, while the current data
are not yet sensitive to the hypothesis of $\tilde{k}=0.1$.

\begin{figure}[hbtp]
  \centering
    \includegraphics[width=1.0\cmsFigWidth]{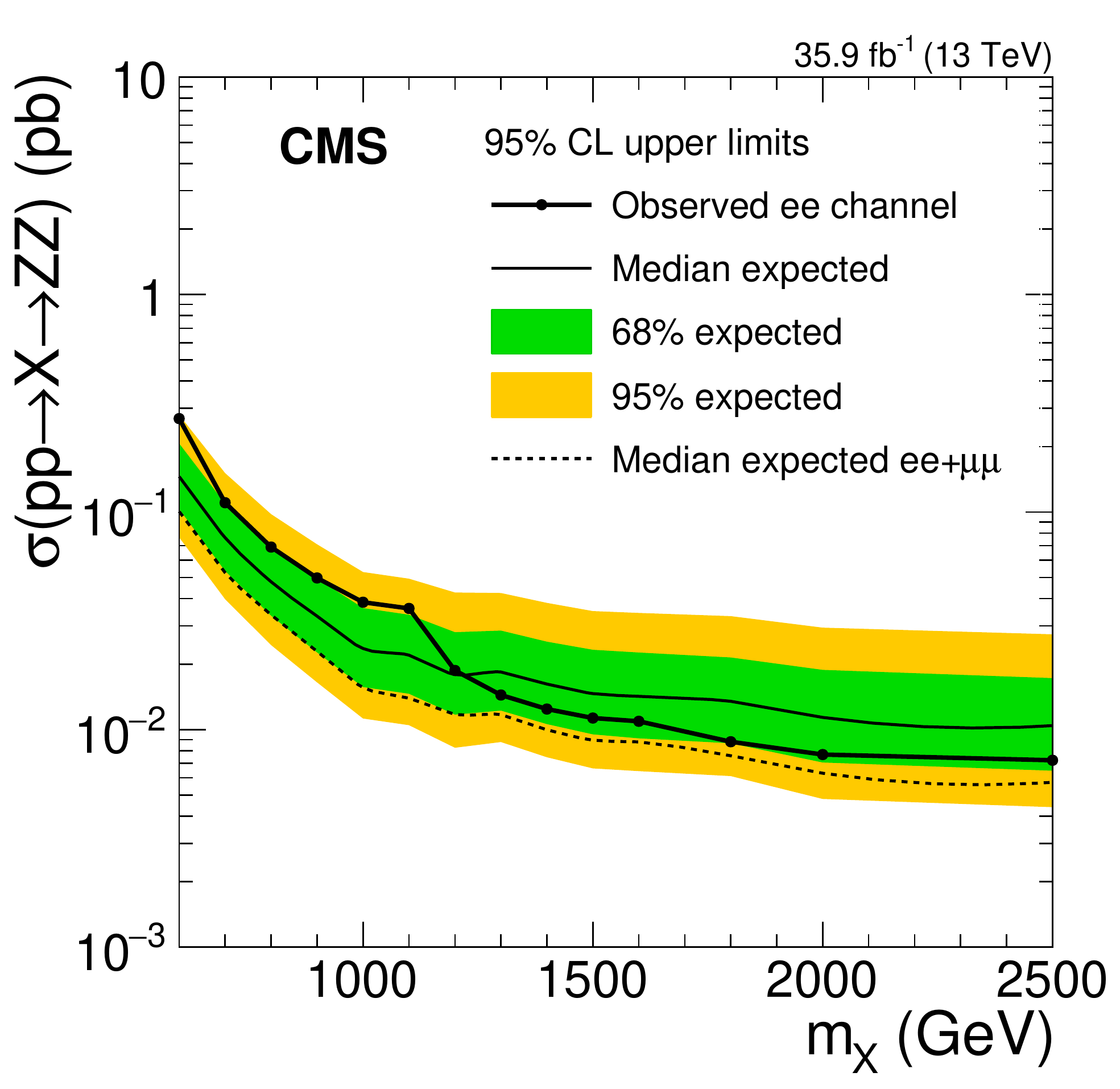}
    \includegraphics[width=1.0\cmsFigWidth]{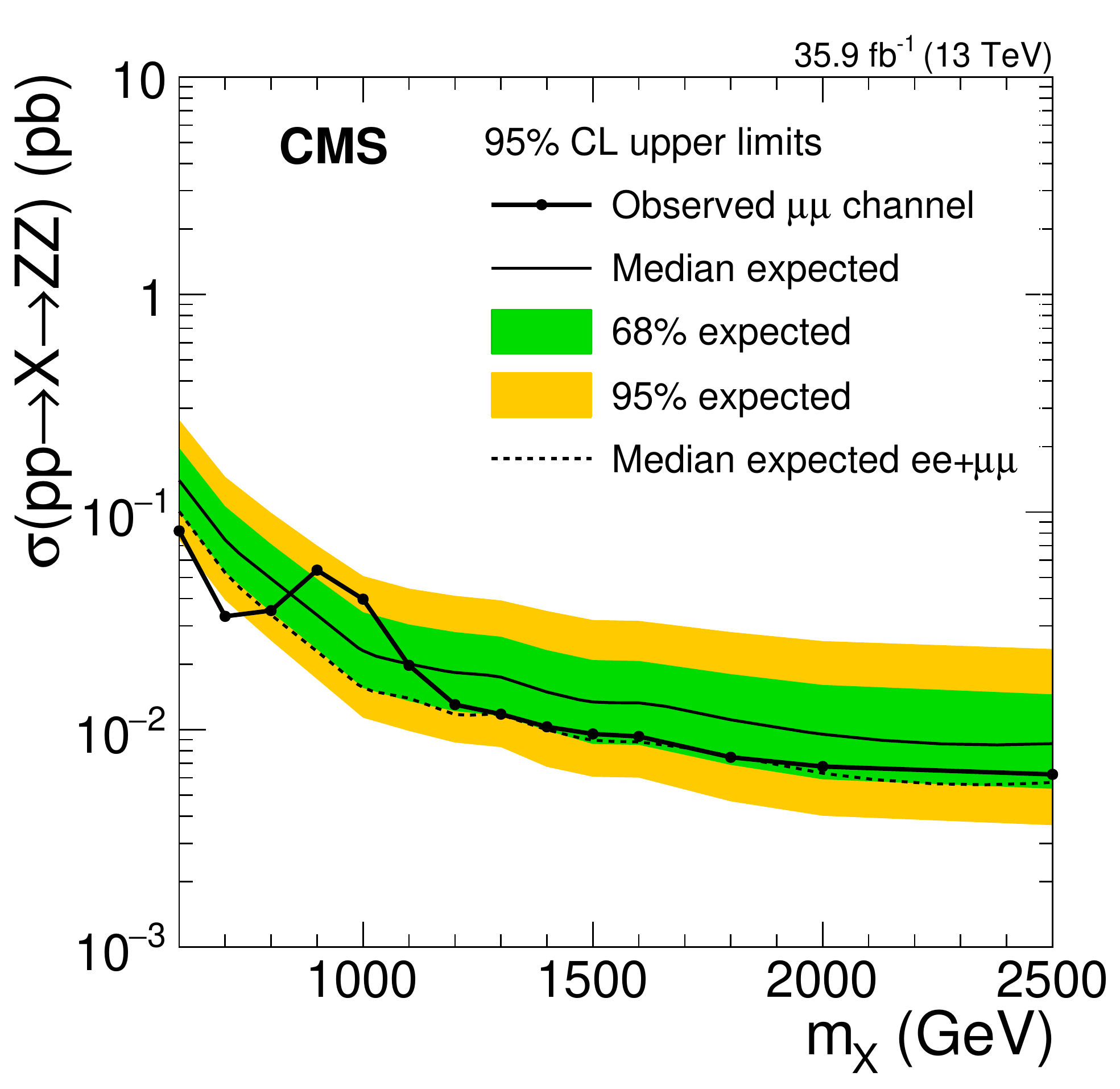}
    \caption{Expected and observed limits on the product of cross section
        and branching fraction of a new spin-2 bulk heavy resonance
        $\resX \to \cPZ\cPZ$, assuming zero width, shown separately for
        searches $\resX\to\cPZ\cPZ\to\ell\ell\Pgn\Pgn$
        in the electron (left) and muon (right) final states.  The median
        expected 95\% CL limits from the combined analysis
        (Fig.~\ref{fig:limits_narrow_sr}) are also shown.
        }
    \label{fig:limits_narrow_sr_ee_mm}
\end{figure}

The observed limits are within 2 standard deviations of expectations
from the background-only model.
The largest upward fluctuations in the data are observed for
$\mX \approx 900$\GeV and
weaken the corresponding exclusions
in this region.
To explore this region in more detail, upper limits are shown
separately for the electron and muon channels in
Fig.~\ref{fig:limits_narrow_sr_ee_mm}.  The upward fluctuations
at $\mX\approx 900$\GeV appear mainly in the muon channel, and
additional fluctuations below this \mX\ can also be observed.

\begin{figure}[hbtp]
  \centering
    \includegraphics[width=1.0\cmsFigWidth]{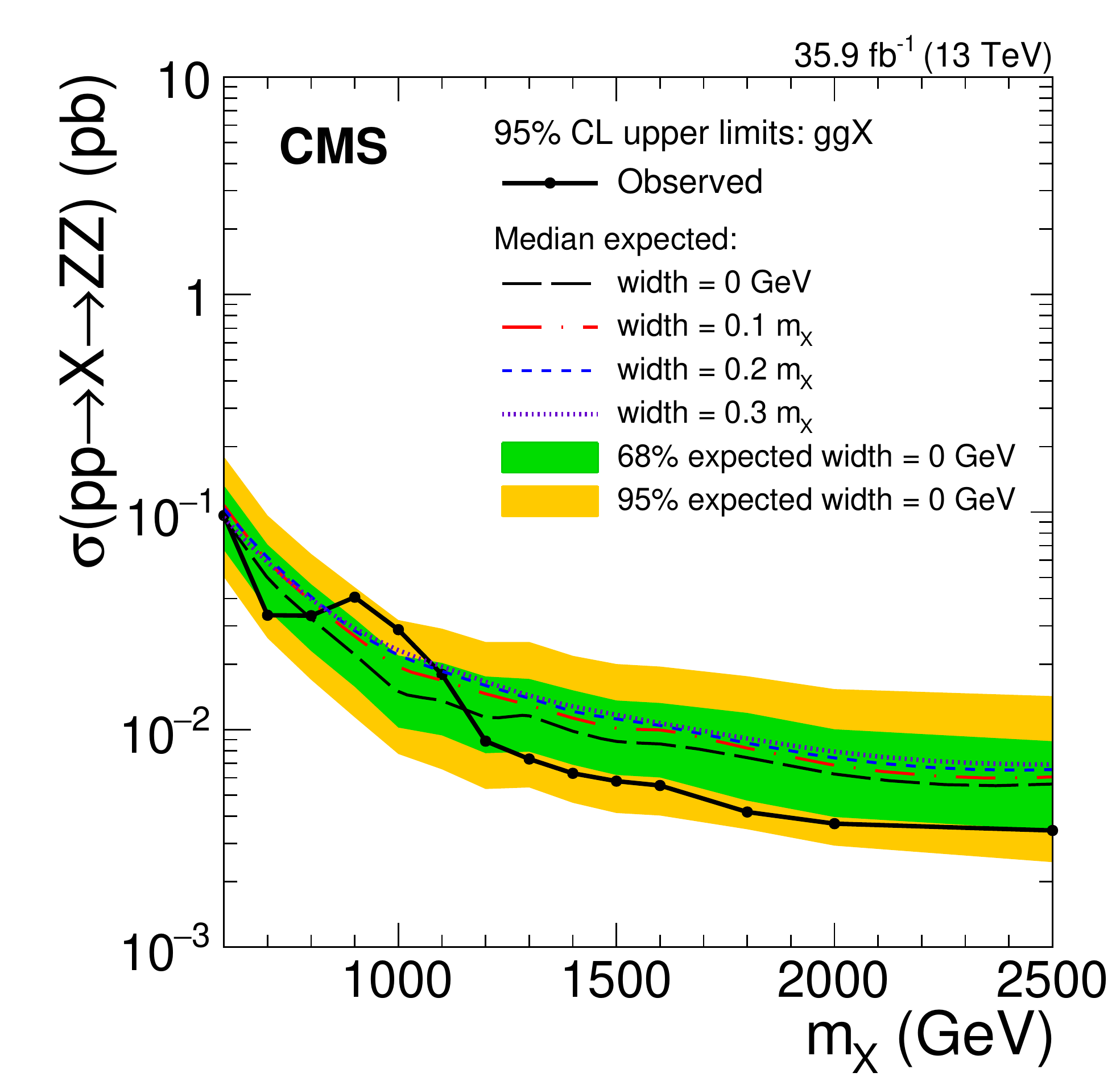}
    \includegraphics[width=1.0\cmsFigWidth]{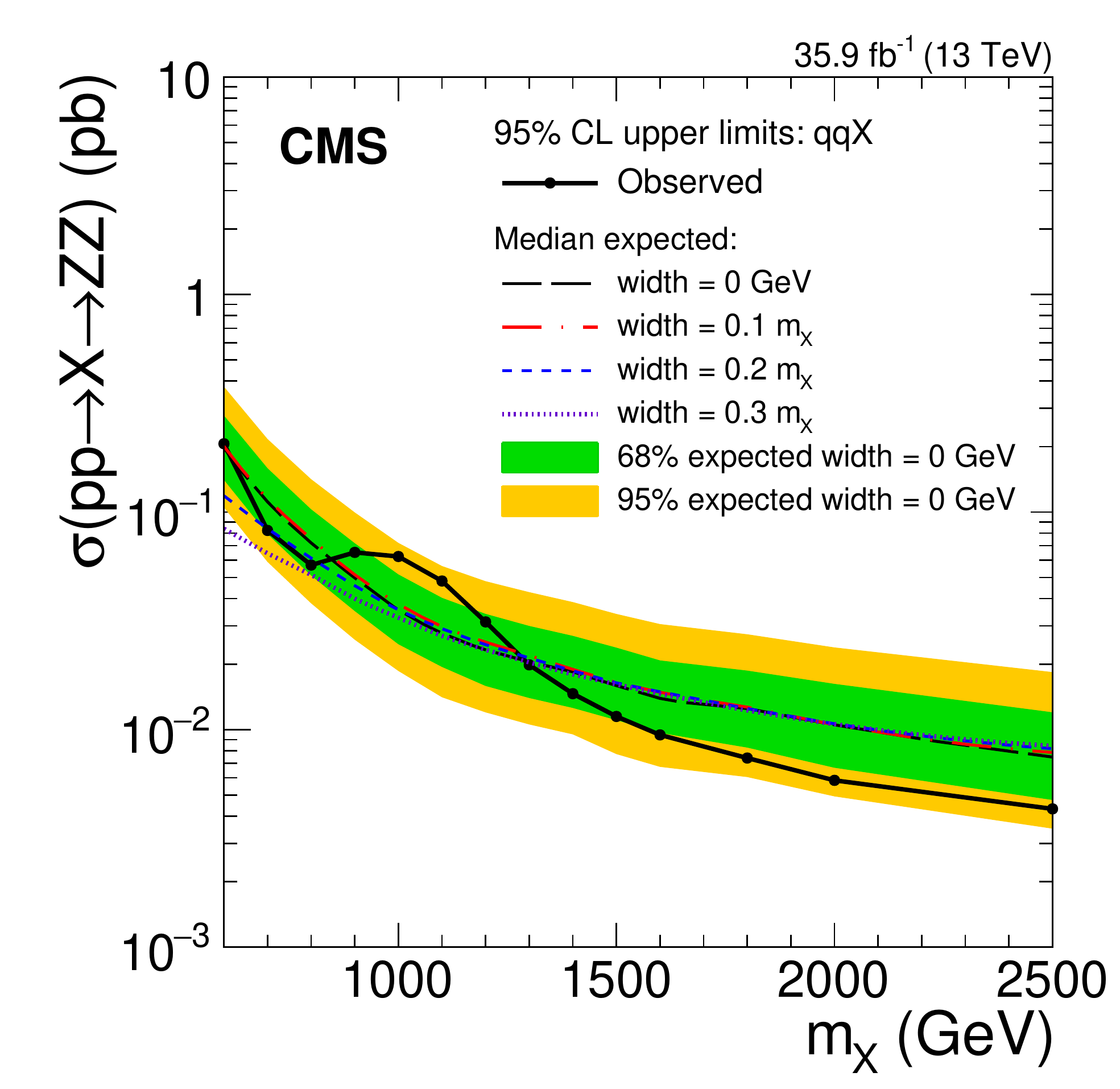}
        \caption{Expected and observed limits on the product of cross section and
        branching fraction of a new spin-2 heavy resonance
        $\resX \to \cPZ\cPZ$ based on a combined analysis of the
        electron and muon channels.  The more generic version of the
        bulk graviton model is considered, assuming either
        gluon-gluon fusion (left) or \qqbar annihilation (right) processes.
        Expected limits are also shown for models having various decay widths
        relative to the mass of the resonance.}
    \label{fig:limits_wide_sr}
\end{figure}

The analysis is repeated comparing to the more general wide width
version of the bulk graviton model described above.
The initial state is fixed purely to either a gluon--gluon fusion or \qqbar
annihilation process and the width of the resonance varied between 0
and $0.3\mX$.  The 95\% CL limits for these models are
shown in Fig.~\ref{fig:limits_wide_sr}.
Differences in the limits between the gluon fusion and \qqbar production
processes arise from spin and parity effects,
which broaden the \mT peak in \qqbar production~\cite{Bolognesi:2012mm}.

\section{Summary}

A search for the production of new resonances has been performed
in events with a leptonically decaying \cPZ{} boson and missing transverse momentum,
using data corresponding to an integrated luminosity
of 35.9\fbinv of proton-proton collisions at a
center-of-mass energy of 13\TeV.
The data are consistent with expectations from standard model
processes.  The hypothesis of a spin-2 bulk graviton, \resX,
decaying to a pair of \cPZ{} bosons
is examined for $600\le \mX\le 2500$\,GeV, and upper limits are set
at 95\% confidence level on the product of the cross section and
branching fraction $\sigma(\Pp\Pp \to \resX\to \cPZ\cPZ)$ ranging from
100 to 4\unit{fb}.  For bulk graviton models characterized by a
curvature scale parameter $\tilde{k}=0.5$ in the extra dimension,
the region $\mX < 800$\GeV is excluded, providing the most
stringent limit reported to date.
The analysis is repeated considering variations of the
bulk graviton model to include a large mass-dependent width.  Exclusion
limits are provided separately for gluon--gluon fusion and \qqbar
annihilation production processes.

\begin{acknowledgments}
We congratulate our colleagues in the CERN accelerator departments for the excellent performance of the LHC and thank the technical and administrative staffs at CERN and at other CMS institutes for their contributions to the success of the CMS effort. In addition, we gratefully acknowledge the computing centers and personnel of the Worldwide LHC Computing Grid for delivering so effectively the computing infrastructure essential to our analyses. Finally, we acknowledge the enduring support for the construction and operation of the LHC and the CMS detector provided by the following funding agencies: BMWFW and FWF (Austria); FNRS and FWO (Belgium); CNPq, CAPES, FAPERJ, and FAPESP (Brazil); MES (Bulgaria); CERN; CAS, MoST, and NSFC (China); COLCIENCIAS (Colombia); MSES and CSF (Croatia); RPF (Cyprus); SENESCYT (Ecuador); MoER, ERC IUT, and ERDF (Estonia); Academy of Finland, MEC, and HIP (Finland); CEA and CNRS/IN2P3 (France); BMBF, DFG, and HGF (Germany); GSRT (Greece); OTKA and NIH (Hungary); DAE and DST (India); IPM (Iran); SFI (Ireland); INFN (Italy); MSIP and NRF (Republic of Korea); LAS (Lithuania); MOE and UM (Malaysia); BUAP, CINVESTAV, CONACYT, LNS, SEP, and UASLP-FAI (Mexico); MBIE (New Zealand); PAEC (Pakistan); MSHE and NSC (Poland); FCT (Portugal); JINR (Dubna); MON, RosAtom, RAS, RFBR and RAEP (Russia); MESTD (Serbia); SEIDI, CPAN, PCTI and FEDER (Spain); Swiss Funding Agencies (Switzerland); MST (Taipei); ThEPCenter, IPST, STAR, and NSTDA (Thailand); TUBITAK and TAEK (Turkey); NASU and SFFR (Ukraine); STFC (United Kingdom); DOE and NSF (USA).

\hyphenation{Rachada-pisek} Individuals have received support from the Marie-Curie program and the European Research Council and Horizon 2020 Grant, contract No. 675440 (European Union); the Leventis Foundation; the A. P. Sloan Foundation; the Alexander von Humboldt Foundation; the Belgian Federal Science Policy Office; the Fonds pour la Formation \`a la Recherche dans l'Industrie et dans l'Agriculture (FRIA-Belgium); the Agentschap voor Innovatie door Wetenschap en Technologie (IWT-Belgium); the Ministry of Education, Youth and Sports (MEYS) of the Czech Republic; the Council of Science and Industrial Research, India; the HOMING PLUS program of the Foundation for Polish Science, cofinanced from European Union, Regional Development Fund, the Mobility Plus program of the Ministry of Science and Higher Education, the National Science Center (Poland), contracts Harmonia 2014/14/M/ST2/00428, Opus 2014/13/B/ST2/02543, 2014/15/B/ST2/03998, and 2015/19/B/ST2/02861, Sonata-bis 2012/07/E/ST2/01406; the National Priorities Research Program by Qatar National Research Fund; the Programa Severo Ochoa del Principado de Asturias; the Thalis and Aristeia programs cofinanced by EU-ESF and the Greek NSRF; the Rachadapisek Sompot Fund for Postdoctoral Fellowship, Chulalongkorn University and the Chulalongkorn Academic into Its 2nd Century Project Advancement Project (Thailand); the Welch Foundation, contract C-1845; and the Weston Havens Foundation (USA). \end{acknowledgments}
\bibliography{auto_generated}

\cleardoublepage \appendix\section{The CMS Collaboration \label{app:collab}}\begin{sloppypar}\hyphenpenalty=5000\widowpenalty=500\clubpenalty=5000\textbf{Yerevan Physics Institute,  Yerevan,  Armenia}\\*[0pt]
A.M.~Sirunyan, A.~Tumasyan
\vskip\cmsinstskip
\textbf{Institut f\"{u}r Hochenergiephysik,  Wien,  Austria}\\*[0pt]
W.~Adam, F.~Ambrogi, E.~Asilar, T.~Bergauer, J.~Brandstetter, E.~Brondolin, M.~Dragicevic, J.~Er\"{o}, A.~Escalante Del Valle, M.~Flechl, M.~Friedl, R.~Fr\"{u}hwirth\cmsAuthorMark{1}, V.M.~Ghete, J.~Grossmann, J.~Hrubec, M.~Jeitler\cmsAuthorMark{1}, A.~K\"{o}nig, N.~Krammer, I.~Kr\"{a}tschmer, D.~Liko, T.~Madlener, I.~Mikulec, E.~Pree, N.~Rad, H.~Rohringer, J.~Schieck\cmsAuthorMark{1}, R.~Sch\"{o}fbeck, M.~Spanring, D.~Spitzbart, A.~Taurok, W.~Waltenberger, J.~Wittmann, C.-E.~Wulz\cmsAuthorMark{1}, M.~Zarucki
% ...
\vskip\cmsinstskip
% ...
1:~~Also at Vienna University of Technology, Vienna, Austria\\
% ...

\end{sloppypar}
\end{document}